\input{aipcheck}

\documentclass[
    ,final            
  ]
  {aipproc}

\layoutstyle{6x9}

\newcommand{\beq}{\begin{equation}}
\newcommand{\eeq}{\end{equation}}
\newcommand{\ba}{\begin{array}}
\newcommand{\ea}{\end{array}}
\newcommand{\beqa}{\begin{eqnarray}}
\newcommand{\eeqa}{\end{eqnarray}}
\newcommand{\bd}[1]{ \mbox{\boldmath $#1$}  }
\newcommand{\oh}{\frac{1}{2}}

\newcommand{\oq}{\frac{1}{4}}
\newcommand{\alp}{\alpha}

\newcommand{\nn}{\nonumber}

\newcommand{\APNY}[1]{Ann. Phys.(N.Y.){\bf {#1}}}

\begin{document}

\title{The Quantum Liquid of Alpha Clusters\\
-a variational approach}


\author{Florin Carstoiu and \c Serban Mi\c sicu}{
  address={NIPNE-HH, Department of Theoretical Physics, 
Bucharest-Magurele, POB MG-6, Romania},
email={misicu@theory.nipne.ro}
}

\begin{abstract}
Within the variational approach of Bose liquids we analyze the g.s. energy of
charge neutral alpha matter at $T$=0. As a prerequisite for such  
calculation we take from the literature or propose new $\alp-\alp$ potentials
that are particularly suitable for this task, i.e. posses a repulsive core 
and/or reproduce the low energy scattering data and the resonance properties of
the $\alp-\alp$ system. The alpha matter EOS is then
obtained with  the HNC method using Pandharipande-Bethe correlation 
derived variationally in the lowest order expansion of the energy functional
or a simple gaussian function with a healing range determined  by the
normalization of the radial distribution function in the lowest order. We show 
that saturation is achieved only via repulsive and shallow
potentials that are not consistent with the scattering and resonance constraints.  
\end{abstract}

\maketitle


\section{Introduction}

The pioneering alpha-matter calculations reported in \cite{clark80} made use 
of $\alp-\alp$ potentials characterized by a strong or even 
infinite repulsive component. These interactions were constructed to fit the 
elastic-scattering phase shifts deduced from experiment. The Equation of State 
(EOS) depends strongly on the shape and strength of these potentials and it is usually derived 
within the frame of the paired-phonon analysis (PPA) or the HNC/n method. The EOS 
calculated with hard-core potentials, trivially saturates at densities  and energies close 
to the nuclear matter saturation point ($\rho_\alp\approx$ 0.04 fm$^{-3}$, 
$E/N_\alp\approx$ -11-16 MeV).
On the other hand, for a typical soft-core potential such as the one proposed by Ali 
and Bodmer (AB) \cite{ali66} the alpha matter almost fails to saturate 
\cite{clark80}. In fact a deep minimum in a very soft EOS at a high density is 
predicted with AB potential. However at such high densities the alpha-condensate 
is almost completely depleted due to Pauli blocking. Somehow this 
disappointing result is conflicting with what one would expect based on the manifestation of alpha clustering in real 
nuclei.
The clusterization of alpha particles on the surface of nuclei at densities
around half the central nuclear density, as revealed by $\alpha$-decay, 
$\alpha$-transfer reactions or the putative dilute three-alphas condensate in the
Hoyle state of $^{12}$C, are pointing to a high stability of alpha matter at
lower densities. 

We recently addressed the problem of alpha matter saturation and concluded
that in order to avoid the collapse of the EOS very shallow potentials with a 
strong repulsive core are necessary \cite{miscar09}. It would be then of interest to extent this study to other potentials, especially those in agreement to elastic
phase-shift data or the first resonant states in the short lived $^8$Be.  
It is therefore timely to gain a better understanding of the role
played by the $\alp-\alp$ potentials in a many-body approach to $\alp$-matter.  

\section{Alpha-Alpha Potentials}

The popular $S$-state Ali-Bodmer $\alpha-\alpha$ potential consists of a 
short-ranged (1.43 fm) repulsive part and a long-ranged attractive 
part (2.50 fm) \cite{ali66},
\beq
V_{\alp\alp}(r)=475\exp\left [{-(0.7r)^2}\right ]
-130\exp\left [-(0.475r)^2\right ].
\eeq
This potential reproduce the $\alp-\alp$ elastic scattering phase-shifts
at low energies. 
The same phase-shift properties are shared by the deep Buck et al. (BFW) 
potential \cite{BFW77} ,
\beq
  V_{\alp\alp}(r)=V_0\exp\left [{-(\mu r)^2}\right ],      
\eeq
where $V_0$ =-122.6225 MeV, $\mu$ = 0.469 fm$^{-1}$. 
In addition the BFW potential reproduces the energy and the 
width of the first $0^+$ resonant state in $^8$Be.

A microscopic approach to the $\alp-\alp$ potential is provided by the 
double-folding method \cite{carst93} where the interaction between two
alpha ions is calculated as an overlap of the single particle densities of the
interacting objects smeared by the  local two-body potential $v_{nn}$, 
\beq
V_{\alp\alp}(\bd{r})=\int d\bd{r}_1 \int d\bd{r}_2
\rho_{\alp}(\bd{r}_1)\rho_{\alp}(\bd{r}_2)
v_{nn}(\rho,\bd{r}-\bd{r}_1+\bd{r}_2)
\label{dfold}
\eeq
The effective  $n-n$ interaction $v_{nn}$ is taken to be dependent
on the density $\rho$ of the nuclear matter where the two nucleons are embedded.
It should also consist of a density independent finite-range part with
preferably two ranges such that a potential similar to the Ali-Bodmer is
obtained at least for the direct component.
A choice satisfying these requirement is provided by the Gogny \cite{Gogny73}
interaction.
Very recently we proposed a bare $\alp-\alp$ interaction based on the 
double-folding method at energies around the barrier, using realistic 
densities of the $\alp$-particle and the density dependent Gogny  
nucleon-nucleon effective interaction \cite{miscar09}. The bare 
$\alpha-\alpha$ potential consisted only in the direct term. In the present work
we add the knock-on nucleon exchange term including recoil corrections and the 
non-local kernels are localized in the lowest order
of the Perey-Saxon approximation at energies around the barrier \cite{carslass96}.
Since all three parametrizations of the Gogny force (D1 \cite{DG80}, D1S \cite{bgg91}, 
D1N \cite{CGS08}) 
are dominated by exchange the resulted folded potentials are very deep.
It is a matter of evidence that deep potentials such as BFW and Gogny
are never saturating alpha matter because of the large
amount of attraction that will contribute to the total energy.


For our alpha matter investigations let us first consider the 
direct part in the double-folding potential (\ref{dfold}). The 
motivation behind such a simplification was recently discussed in connection 
with the necessity of accounting for the incompressibility of nuclear matter in 
cold clustering processes \cite{del01}  and  extreme sub-barrier fusion 
\cite{misesb06}. In this framework a double-folding repulsive potential with a 
zero-range interaction is added to the direct and exchange potential such that the
energy cost for overlapping two pieces of nuclear matter are payed-off.
The strength of this repulsive $\delta$-like potential is in a simplified
picture proportional to the nuclear incompressibility at the corresponding
density of total overlap.

We assume a Gaussian nuclear matter distribution of the $\alpha$-particle,
\beq
\rho_\alp(r)=4\left ( \frac{1}{\pi b^2}\right )^{3/2} e^{-r^2/b^2},
\label{rhoalf}
\eeq
where the length parameter $b$ is determined from the root mean square radius (rms)
1.58$\pm$0.002 fm extracted via a Glauber analysis of experimental interaction
cross sections \cite{AlK96}.

Using the analytical expression of the direct component $n-n$ force in the 
Gogny parametrization, as listed in  \cite{miscar09},  and inserting the Gaussian 
density distribution (\ref{rhoalf}) in the double folding integral (\ref{dfold})
the direct part of the $\alp-\alp$ interactions reads,
\beqa
V_{\alp\alp}^{\rm d}(r)&=&4\sum_{i=1}^2(4W_i+2 B_i-2H_i-M_i)\left
(\frac{\mu_i^2}{\mu_i^2+2b^2}\right )^{3/2}e^{-\frac{1}{\mu_i^2+2b^2}{r^2}}\nn\\
&+&\frac{3}{2}t_3\frac{4^{\gamma+2}}{(\gamma+2)^{3/2}(\sqrt{\pi}b)^{3(\gamma+1)}}
e^{-\frac{\gamma+2}{4b^2}{r^2}}
\eeqa
Above, $W_i,B_i,H_i,M_i$ are strength of the $i=1,2$-th finite-range term 
of the Gogny parametrization whereas $\mu_i$ the corresponding ranges, while
the parameter $\gamma$ characterizes the density dependence of the force.
\begin{figure}[h]
\includegraphics[height=.32\textheight]{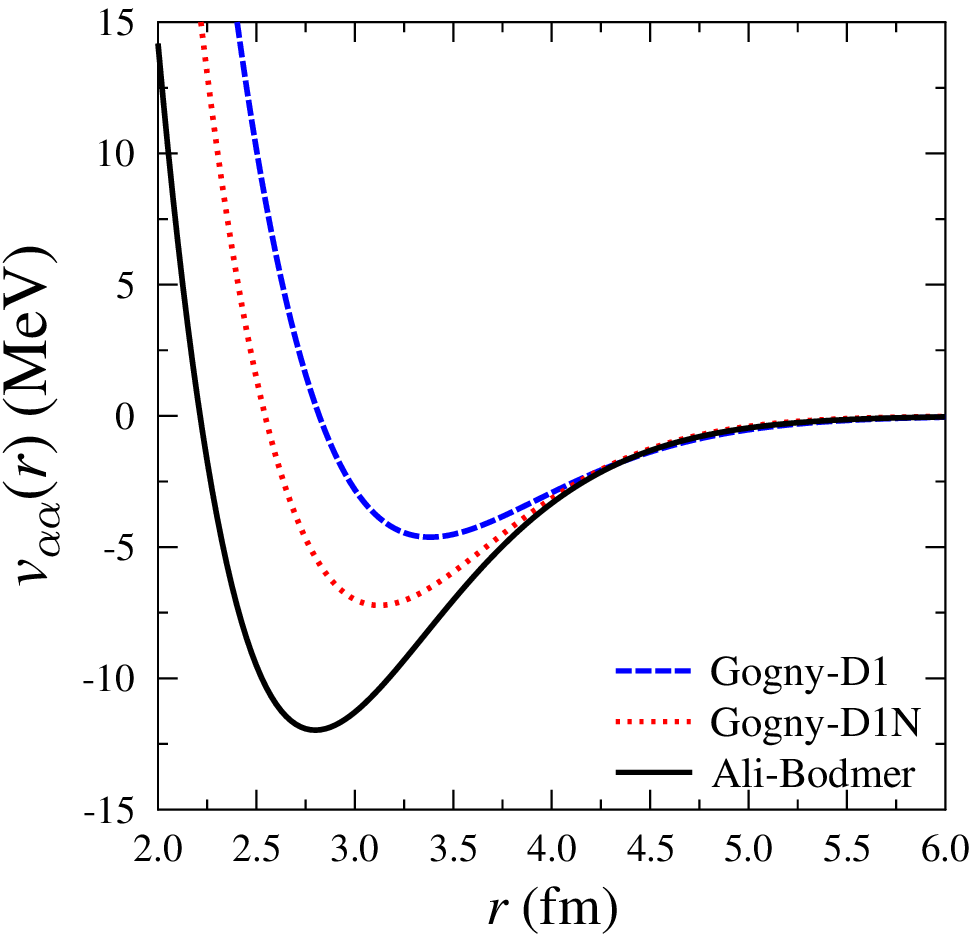}
\includegraphics[height=.32\textheight]{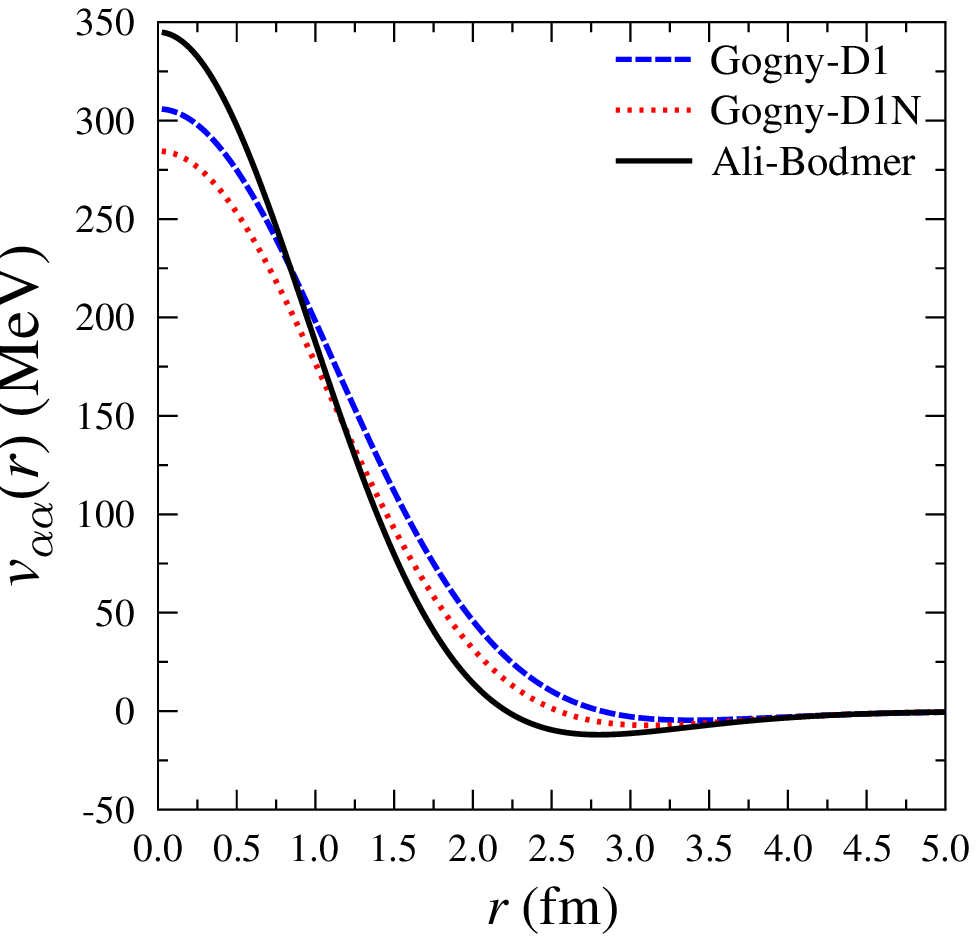}
\caption{ Comparison of the Gogny D1 and D1N direct $\alp-\alp$ potentials with
the AB potential.}
\label{Figpot1}
\end{figure}
In the left panel of Fig.\ref{Figpot1} we represent the AB, D1 and D1N potentials on a magnified scale around the minimum. The two Gogny potentials  display pockets that are shallower and shifted  to larger radii compared to the Ali-Bodmer potential. The minima of these potentials are close to the  classically 
touching configuration of the $\alp-\alp$ system.


The derivation of the exchange part of the 
potential will be given elsewhere. Here we only reproduce the 
closed expression obtained after the localization via the Perey-Saxon method
of the non-local kernel, 
\beqa
V_{\alp\alp}^{\rm ex}(r)&=&-32\sum_{i=1}^2(W_i+2 B_i-2H_i-4M_i)
\left (\frac{\beta_i}{b}\right )^3 e^{-\frac{1}{2b^2}
\left[1-\oq\left (\frac{\beta_i}{b}\right )^2\right ]r^2}
e^{\pm\oh |K|^2\beta_i^2}\nn\\
&\times&
\left\lbrace
\begin{array}{ccc}
\exp{\left [-\oh \left (\frac{\beta_i}{b}\right )^2 {K}{r}\right ]} 
& {\rm for} & K^2 < 0\\
\cos\left [ \oh\left (\frac{\beta_i}{b}\right )^2|{K}|{r}\right ]
& {\rm for} & K^2 \geq 0
\end{array}
\right. 
\eeqa
where,
\beq
\frac{1}{\beta_i^2}= \frac{8}{\mu_i^2}+\frac{9+\oq}{b^2}.
\eeq
and,
\beq
K^2(r)=\frac{2\mu}{\hbar^2}(E_{c.m.}-V_{\alp\alp}(r))
\eeq
Above we distinguish between  a sub-barrier branch of the potential ($K^2<$0) and 
an over-barrier one ($K^2>$0).
Note that in the exchange kernel we used an alpha single-particle density matrix
constructed from the 0$s$ harmonic-oscillator orbitals.
Only a few iterations are needed to obtain convergence in the localization
procedure. In the left panel of Fig.\ref{Figpot2} the full (direct+exchange) 
$\alp-\alp$ potential for three parametrizations of the 
Gogny interaction is displayed 
at zero relative momentum ($K_{\rm rel}=0$) and compared to the BFW potential.

\subsection{Supersymmetric Potentials}

In a previous work we concluded that a shallow 
potential (in any case shallower than the traditional AB) is 
needed in order to obtain the saturation alpha matter 
at physically reasonable densities \cite{miscar09}.
On the other hand, a shallow potential results naturally from the 
requirement that $^8$Be has no bound states.
An alternative way to the Ali-Bodmer phenomenological ansatz to 
construct a shallow potential was proposed by Baye \cite{baye87}. 
A sequence of supersymmetric transformations, which removes the Pauli-forbidden 
states in the sense of orthogonality condition model (OCM), is applied to the 
original Schr\"odinger equation in such a way that the phase-shifts
corresponding to the transformed Hamiltonian coincide to the
phase-shifts of the original Hamiltonian. In this manner one
obtains a new shallow potential that is exactly phase-equivalent 
(modulo $2\pi$) to the original deep potential. In the particular case of $\alp-\alp$
system the phase-equivalent potential is obtained from the
suppression of the two lowest ($l$=0) bound forbidden states. A compact formula
provides the needed shallow potential 
\beq
V_{\alp\alp}^{(2)}(r)=V_{\alp\alp}^{(0)}(r)-\frac{\hbar^2}{m}
\frac{d^2}{dr^2}{{log}}\det{\bd \Psi}^{(2)}(r)
\eeq
where the elements of the 2$\times$2 matrix ${\bd \Psi}^{(2)}$ read
\beq
({\bd \Psi}^{(2)})_{ij}(r)=\int_0^r u_i(r')u_j(r')dr'~~ (i,j=0s,1s)
\eeq
and $V_{\alp\alp}^{(0)}$ is any one of the deep potentials
discussed earlier.
Above we denote by $u_{0s,1s}$ the lowest two bound state wave functions 
of the deep nuclear+Coulomb ($l$=0) potential. 
We obtain these states using the Runge-Kutta-Nystr\"om (RKN) numerical 
integration \cite{ixar80}.

The suppersymmetric shallow partners of the BFW potential were 
determined already by Baye \cite{baye87} and more recently in 
a semi-classical approach in \cite{hbi08} for $l$=0, 2 
and 4. The first two bound states in this potential are located 
according to this reference at -72.8 MeV and -25.9 MeV.
With the RKN method we obtain the values -72.62 MeV and -25.61 MeV.
The minimum of the shallow BFW-SUSY potential is located at 2.85 fm and has 
a depth of -7.56 MeV and is similar to the Baye estimation and about 1 MeV 
shallower than the semiclassical estimation of Horiuchi et al. \cite{hbi08}.



The transformed potential displays a $20/r^2$ singularity at the
origin. In practical calculations we use an interpolation with 
gaussians that is non-singular at the origin. 

\begin{figure}
\includegraphics[height=.32\textheight]{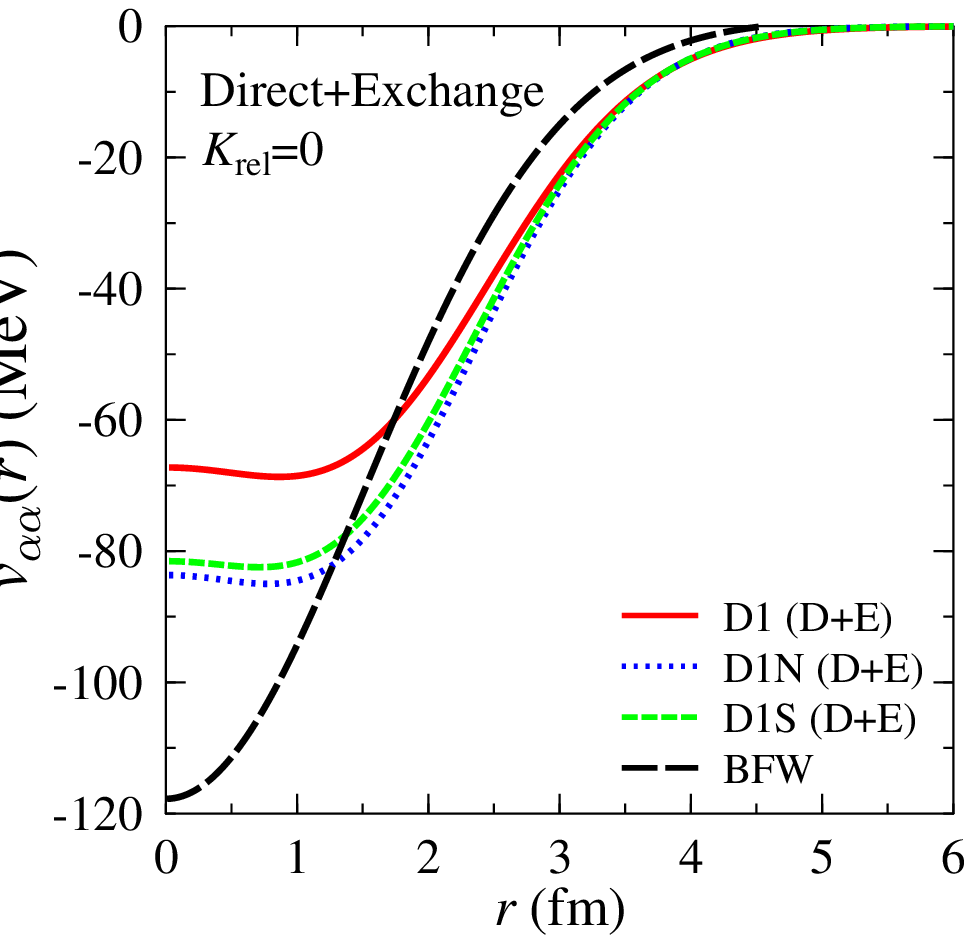}
\includegraphics[height=.32\textheight]{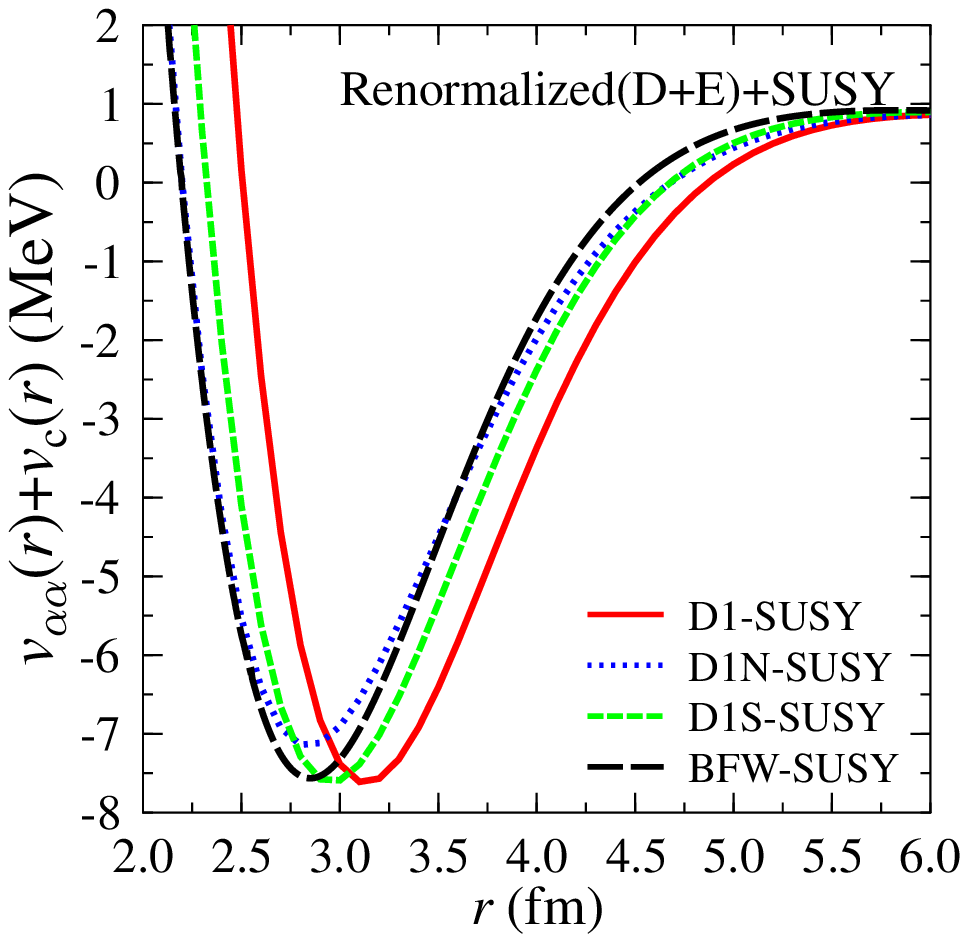}
\caption{Bare direct+exchange  $\alpha-\alpha$ potentials for the BFW and the 
three $n-n$ Gogny interactions used in the text (left panel) and their 
supersymmetric partners (right panel).}
        \label{Figpot2}
\end{figure}

To obtain the SUSY partner of the Gogny potentials we constrain first 
the original deep potential to reproduce the properties of
the first resonant state in $^8$Be, i.e. 
$E_{\rm res}=$92.12 $\pm$ 0.05 keV, $\Gamma_0$=6.8$\pm$ 1.7 eV \cite{benn66}.
The renormalized Gogny potentials are
next subjected to a sequence of two supersymmetric transformations in order
to eliminate the Pauli-forbidden states. 
In the right panel of Fig.\ref{Figpot2} we display the supersymmetric 
partners of the BFW and the three Gogny interactions (including the Coulomb 
component). We make the remark that the deep potentials BFW, D1 and D1S(D1N)
have very different depths but upon renormalization and supersymmetric transformations all four 
potentials have similar depths. Thus, the physical contraint imposed by
the $0^+$ resonance in $^8$Be and the removal of the forbidden states,  lead to an 
almost unique potential for the $\alp-\alp$ system.


\section{EOS of Alpha Matter}

\begin{figure}
\includegraphics[height=.36\textheight]{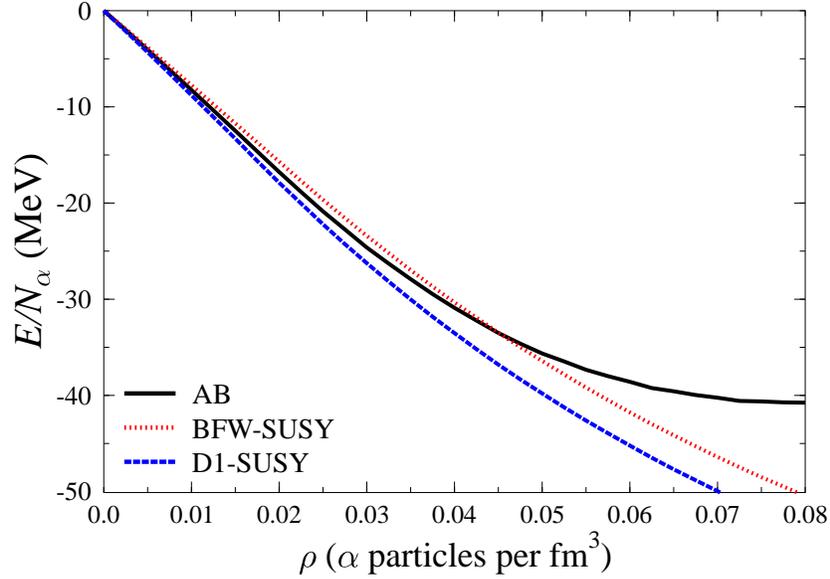}
\caption{Comparison between the EOS predicted with HNC/0 for
AB, BFW-SUSY and D1-SUSY.}
\label{eos_hnc0_susy}
\end{figure}

We derive the EOS of alpha matter using the variational approach,
described in \cite{miscar09}. In Fig.\ref{eos_hnc0_susy} we compare 
the EOS for AB potential with  BFW-SUSY and D1-SUSY and we conclude that
for this last two cases the EOS curves are collapsing, while the EOS for 
the AB interaction saturates at too large densities. Thus shallow potentials 
constrained to reproduce the phase-shift elastic data and/or the first resonant 
state in $^8$Be do not saturate alpha matter in the HNC/0 approximation. 
The reason for this disappointing result is that the two-body correlation
functions obtained with the Pandharipande-Bethe prescription display large
overshootings near the healing distance with increasing alpha matter density.
The density dependence of the healing distance $d$ does not bring sufficient
density dependence in order to ensure saturation.

\begin{figure}[t]
\includegraphics[height=.7\textheight]{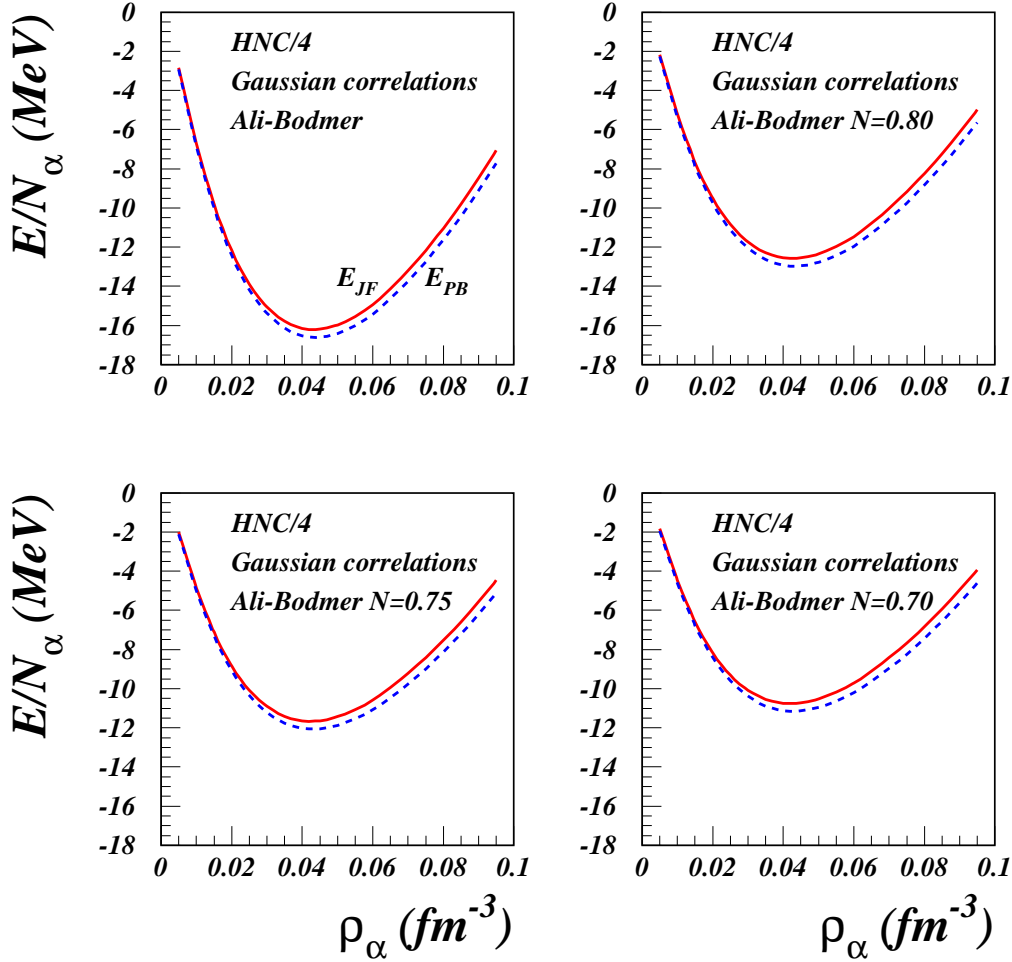}
\caption{Equation of state for alpha matter in the HNC/4 approximation. The bare
AB potential is scaled with a factor $N$ in order to see the variation of the
saturation point as a function of the potential strength.}
        \label{fighnc4}
\end{figure}

In order to demonstrate this effect we consider a simple phenomenological
two-body Gaussian correlation function \cite{bala},
\beq
f(r)=1-e^{-\beta^2r^2}
\eeq
Evidently this function has no overshooting. The parameter $\beta$ is determined
from the normalization condition \cite{feenberg},
\beq
4\pi\rho\int_0^{\infty} dr ~r^2(f^2(r)-1)=-1
\label{eqnorm}
\eeq
where $\rho$ is the density of alpha matter. We perform a 
selfconsistent HNC/4 calculation
using these correlation functions and the bare AB potential. Both Jackson-
Feenberg ($E_{JF}$) and Pandharipande-Bethe ($E_{PB}$) are displayed in
Fig.\ref{fighnc4} for different renormalizations $N$ of the AB
potential. The intrinsic binding energy of the $\alp$ particle ($B_\alp\sim$ 7 MeV)
is not included.
The PB energy is calculated using the Kirkwood approximation for the three-body
radial distribution. The values obtained in this approximation are quite
similar with those obtained with JF approximation since the contribution from
the three-body term is negligibly small. The saturation density does not change
with the renormalization of the potential since this is determined exclusively
by the density dependence of the range parameter $\beta\sim\rho^{1/3}$
(Eq.\ref{eqnorm}). Similar results have been obtained for the equation of state 
using a 4$^{th}$ order cluster expansion and the HNC/0 method.
Regardless of the normalization factor the EOS saturation point is very close
to the saturation point of normal nuclear matter, 
i.e. $\rho_{\rm NM}=4\rho_{\alp}\approx$ 0.16 fm$^{-3}$.

In conclusion, shallow phenomenological and semimicroscopic $\alp-\alp$ potentials 
which reproduce  low energy $\alp-\alp$ scattering phase shifts and/or the resonance in $^8$Be lead to a collapsing alpha matter equation of state  if two-body correlation functions are obtained with the Pandharipande-Bethe prescription. 
In this prescription, the assumption of a finite weakly density dependent healing 
distance leads to overshooted two-body correlation functions and enhances the importance of the long attractive tail of the
potential.





\begin{theacknowledgments}
This work received  support from CNCSIS Romania,
under Programme PN-II-PCE-2007-1, contracts No.49 and No.258.
\end{theacknowledgments}



\bibliographystyle{aipproc}   


\IfFileExists{\jobname.bbl}{}
 {\typeout{}
  \typeout{******************************************}
  \typeout{** Please run "bibtex \jobname" to optain}
  \typeout{** the bibliography and then re-run LaTeX}
  \typeout{** twice to fix the references!}
  \typeout{******************************************}
  \typeout{}
 }

\end{document}